\documentclass[pra,10pt,aps,reprint,amsmath,amssymb,floatfix,superscriptaddress,longbibliography]{revtex4-1}

\usepackage[utf8]{inputenc}
\usepackage{graphicx}
\usepackage{epstopdf}
\usepackage{amsthm}
\usepackage{empheq}
\usepackage{bbm}
\usepackage{braket}

\newcommand{\id}{\mathbbm{1}}                                          
\newcommand{\tr}[1]{\operatorname{\textnormal{Tr}}\left[ {#1} \right]} 

\begin{document}

\title{Past observable dynamics of a continuously monitored qubit}
\author{Luis Pedro Garc\'ia-Pintos}
\affiliation{Institute for Quantum Studies, Chapman University, Orange, CA 92866, USA}
\author{Justin Dressel}
\affiliation{Institute for Quantum Studies, Chapman University, Orange, CA 92866, USA}
\affiliation{Schmid College of Science and Technology, Chapman University, Orange, CA 92866, USA}

\date{\today}

\begin{abstract}
Monitoring a quantum observable continuously in time produces a stochastic measurement record that noisily tracks the observable. For a classical process such noise may be reduced to recover an average signal by minimizing the mean squared error between the noisy record and a smooth dynamical estimate. We show that for a monitored qubit this usual procedure returns unusual results. While the record seems centered on the expectation value of the observable during causal generation, examining the collected past record reveals that it better approximates a moving-mean Gaussian stochastic process centered at a distinct (smoothed) observable estimate. We show that this shifted mean converges to the real part of a generalized weak value in the time-continuous limit without additional postselection. We verify that this smoothed estimate minimizes the mean squared error even for individual measurement realizations. We go on to show that if a second observable is weakly monitored concurrently, then that second record is consistent with the smoothed estimate of the second observable based solely on the information contained in the first observable record. Moreover, we show that such a smoothed estimate made from incomplete information can still outperform estimates made using full knowledge of the causal quantum state.
\end{abstract}

\maketitle

Over the past decade, time-continuous quantum measurements \cite{Mensky1979,Barchielli1982,Caves1986,Caves1987,Diosi88,Wiseman1993,Mensky1994,Goetsch1994,Korotkov2001} of superconducting qubits (such as transmons \cite{Koch2007}) have become an important and increasingly well-controlled component of emerging quantum computing technology \cite{Katz2006,Palacios-Laloy2010,Vijay2012,Riste2013,Hatridge2013,Murch2013,Ibarcq2013,Weber2014,deLange2014,Roch2014,Campagne-Ibarcq2014,Ibarcq2015,naghiloo2015fluores,Slichter2016,Hacohen-Gourgy2016,chantasri2016,Jordan2016,Ibarcq2016,Foroozani2016,Naghiloo2017,Atalaya2017,Hacohen-Gourgy2017}. Indeed, the primary method for extracting information from a superconducting transmon is to dispersively couple it to a pumped microwave resonator, then amplify and mix the leaked microwave field with a local oscillator to perform a homodyne measurement of the traveling field, which produces a stochastic time-dependent voltage that encodes information about the transmon energy basis \cite{Gambetta2008,Korotkov2014,Korotkov2016}. Understanding what information is contained in the resulting stochastic readout is thus an essential theoretical issue.

In simple terms, a continuous measurement can be understood as a sequence of weak measurements \cite{BookDavies,AharonovWV} on the qubit. In the superconducting case, each temporal segment of the steady-state traveling coherent microwave field acts as an independent and approximately Gaussian meter that becomes entangled with the qubit and later measured \cite{Korotkov2016}. During the measurement of the field, the finite bandwidth of the circuitry typically discretizes the field into time bins of size $dt$. Provided that $dt$ is longer than the correlation timescale of the traveling field, the statistics of the averaged homodyne voltage collected in each independent time bin are approximately Gaussian, producing a discrete temporal sequence of Gaussian-distributed measurement results $\{r_j\}$ with a wide variance that inversely depends upon the time step size $dt$. All information about the qubit must be extracted by processing this stochastic time series. 

For convenience, this time series is traditionally interpolated to construct a time-continuous stochastic process $r(t)$ that preserves the physically correct averages over the time bins $dt$. Such an interpolation then has the structure of a moving-mean stochastic white noise process. The mean of this process is widely recognized to be the expectation value of the monitored observable \cite{Jacobs2006,Wiseman2009}, following straightforward arguments about the increasing width of the Gaussian distributions in the time-continuous limit. This understanding of the mean as an expectation value raises the natural question whether the white noise may be reduced by classical signal processing techniques to recover that expectation value via temporal averaging, rather than ensemble averaging. Indeed, such temporal averaging exposes quantum jumps between measurement eigenstates in the quantum Zeno regime \cite{Vijay2011,Slichter2016,Hacohen-Gourgy2017}. More dramatically, simultaneously monitoring multiple orthogonal observables and processing the collected readouts with simple exponential filtering has been used to estimate an evolving qubit state with surprisingly high fidelity \cite{KorotkovXYZ,LPGPJD16}. One might therefore suspect that optimizing the classical signal processing of the readout could allow quantum state tomography for a single realization of an evolving qubit state to high accuracy with minimal prior information, and thus challenge the operational interpretation of the quantum state as describing only ensemble statistics.

In this paper, we carefully revisit the derivation of the collected readout as a stochastic process and show the counter-intuitive result that the moving mean is \emph{not} in fact the expectation value of the observable, as is usually assumed. Instead, the mean is altered by the measurement backaction away from the expectation value. As such, optimally removing the noise from the readout will not recover the causal quantum state, as might be suspected, which will bound the achievable fidelity of classical filtering state tomography schemes. Instead, the moving mean tracks a \emph{smoothed estimate} that we show converges to the real part of a generalized quantum weak value~\cite{AharonovWV,Jozsa07,DresselWVreview,Dressel2015WVInt} in the time-continuous limit. Notably, this weak value naturally appears without additional postselection for each individual measurement realization. We derive this result analytically, then verify numerically that for individual trajectories the mean squared error of this smoothed estimate is consistently smaller than that of the expectation value. This minimization of the mean squared error is consistent with the usual metric of classical signal processing for determining the optimal estimate of a time-dependent noisy signal. We independently verify the result numerically using a hypothesis testing approach, confirming that the smoothed estimate is indeed a better fit to the collected data than the expectation value. We go on to show that in the presence of a simultaneous second observer, the smoothed estimate retains its objective character. That is, a smoothed estimate made from incomplete data taken only by the first observer can be a better fit to the unknown data of the second observer than even the pure causal qubit state that uses all available data. Notably, this last result improves upon a recent proposal \cite{GuevaraWiseman15} that constructs a ``smoothed quantum state'' to estimate the observations made by an unknown second observer, since that method can never outperform the most pure causal state that uses all collected data. The conclusions of our study are consistent with prior work concerned with time-symmetric quantum state estimates, such as the two-state-vector formalism \cite{Watanabe1955,Aharonov1964,AharonovWV}, quantum smoothing \cite{PRLTsang09,PRATsang09,PRA2Tsang09,RalphWiseman10}, bidirectional quantum states \cite{Dressel13}, and past quantum states \cite{Molmer13,Tan2015,Haroche15,Molmer16}. However, we emphasize here the practical consequence for ongoing research into continuous quantum measurements: \emph{Applying optimal classical signal processing techniques to a single realization of collected data from a continuous quantum measurement produces results that do not correspond to the causal quantum state.}

The paper is organized as follows. 
In Section~\ref{sec:sequencemeasurements} we briefly review the derivation of a simple continuous quantum measurement from a quantum information perspective to recover the usual interpretation of the readout. 
In Section~\ref{sec:quantumsmoothing} we revisit the structure of the past readout given the posterior information about what was collected later in time, showing that the measurement backaction has fundamentally changed its structure. 
In Section~\ref{sec:quantumsmoothingfigures} we consider an explicit example of a Rabi oscillating qubit and compare the observable estimates to the readout in more detail, showing that the smoothed estimate indeed fits the readout better.
In Section~\ref{sec:smoothingphysical} we show that smoothed estimates are objective even in the presence of a second simultaneous observer, thereby reinforcing their operational relevance.  
We conclude in Section~\ref{sec:conclusion}.

\section{Observable dynamics from anterior measurements}
\label{sec:sequencemeasurements}

We focus our discussion on what can be inferred from a collected measurement record about its associated observable dynamics. As such, in what follows we consider a simplified model of time-continuous measurements that is adequate for isolating the relevant features. To keep this manuscript self-contained, we briefly review the essential details of how a temporal sequence of independent Gaussian measurements models continuous-in-time measurements from a quantum information perspective. This model is a slight idealization of those that describe recent experimental work on quantum state trajectories well \cite{Murch2013,Weber2014,Hacohen-Gourgy2016,chantasri2016,Jordan2016,Hacohen-Gourgy2017}, but deliberately neglects relevant experimental details---such as measurement inefficiency, environmental decoherence, energy relaxation, phase-backaction, and non-Markovian effects from finite detector bandwidth \cite{Korotkov2014,Korotkov2016}---in order to isolate the essential effect of the measurement backaction. (For an alternative recent derivation of a simple continuous measurement model that includes some of these nonidealities in the context of feedback, see also Ref.~\cite{Patti2017}.)

Consider a system, such as a qubit, that is assigned a quantum state represented by a density operator $\rho$. To measure an observable $A = \sum_a a\ket{a}\!\bra{a}$ of the system, such as the Pauli operator $\sigma_z$, we couple the system to a measurement device, which reports classical results $r\in\mathbb{R}$ that are correlated with the distinct values of $A$. That is, each distinct value $a$ of $A$ corresponds to a probability measure $P(r|a)\,dr$ for obtaining   $r$ on the detector given that particular $a$, such that each measure is normalized over the possible results of the measurement, $\int_{\mathbb{R}} P(r|a)\,dr = 1$, and the total probability for obtaining a measurable subset $R \subset \mathbb{R}$ of $r$ is $\int_{R}P(r|\rho)\,dr = \int_R\sum_a P(r|a)P(a|\rho)\,dr$ with $P(a|\rho)=\bra{a}\rho\ket{a}$. As a convenient way to formally encapsulate these detector properties, the map $a \to P(r|a)\,dr$ from observable values to probability measures generates a map $A \to E_r\,dr = (\sum_a P(r|a)\ket{a}\!\bra{a})\,dr$ from the observable operator $A$ to a probability operator-valued measure (POVM) $E_r\,dr$. The positive Hermitian operators $E_r$ of the POVM then partition unity $\int_{\mathbb{R}} E_r\, dr = \id$, and obey L\"uder's probability rule \cite{NielsenChuang}, $P(r|\rho)\,dr = \tr{E_r\, \rho}\,dr$.

Such a generalized measurement of $r$ on the detector produces measurement backaction on the state $\rho$ described by a quantum instrument, $\rho \stackrel{r}{\longrightarrow} \mathcal{E}_r(\rho)\,dr$, which is a completely positive map-valued measure satisfying $P(r|\rho)\,dr = \tr{\mathcal{E}_r(\rho)}\,dr$ \cite{Dressel13}. In the case of no classical mixing from information loss, this instrument can be represented by a single $r$-dependent Kraus operator $M_r$ according to $\mathcal{E}_r(\rho)\,dr = M_r \rho M_r^\dagger\, dr$, which relates $M_r$ to the POVM $E_r$ according to $E_r\,dr = M_r^\dag M_r\,dr$. As such, the Kraus operator factors into a polar form $M_r = U_r \sqrt{E_r}$, with $\sqrt{E_r} = \sum_a \sqrt{P(r|a)}\ket{a}\!\bra{a}$ corresponding to partial state collapse (informational backaction) and $U_r$ corresponding to additional $r$-dependent unitary perturbation (stochastic Hamiltonian backaction). In what follows we neglect any unitary backaction $U_r$ for simplicity to focus solely on the effects of the informational backaction. (See Refs.~\cite{Korotkov2014,Korotkov2016,deLange2014} for discussion about the role of such unitary phase-backaction in measurements of superconducting qubits.)

The renormalized state after the observation of a subset $R$ of $r$ values on the detector (e.g., from classically coarse-grained resolution) is 
\begin{align}
\label{eq:updaterhofull}
	\rho \stackrel{R}{\longrightarrow} \frac{\int_R \mathcal{E}_r(\rho)\, dr}{\int_R P(r|\rho)\,dr} = \frac{\int_R M_r \rho M_r^\dag\, dr}{\int_R P(r|\rho)\,dr}.
\end{align}
We will restrict our discussion to a perfect detector with infinitely sharp resolution of individual points $r$ for simplicity, so that the measure factors in Eq.~\eqref{eq:updaterhofull} simply cancel to yield the simplified expression
\begin{align}
\label{eq:updaterho}
	\rho \stackrel{r}{\longrightarrow} \frac{\mathcal{E}_r(\rho)}{P(r|\rho)} = \frac{M_r \rho M_r^\dag}{P(r|\rho)}.
\end{align}

In the following we assume Gaussian measurements of $A = \sigma_z = \ket{1}\!\bra{1} - \ket{0}\!\bra{0}$ describing the computational basis of a qubit. That is, the detector distributions for the distinct values $\pm 1$ of $\sigma_z$ are Gaussian, $P(r|\pm\!1) = G_{\pm 1}(r) \equiv \exp(-(r\mp 1)^2dt/2\tau)/\sqrt{2\pi\tau/dt}$, with equal variances $\tau/dt$ but distinct means centered at their associated values of $\pm 1$. 
This parametrization is chosen such that $dt$ is a \emph{discretization timescale} that specifies the duration of the coupling required to obtain the result $r$, and $\tau$ is a \emph{measurement collapse timescale} that indicates the coupling duration needed to obtain a unit signal-to-noise ratio for the measurement. This variance scaling also guarantees that sequences of independent such measurements correctly average to coarsen the discretization timescale, i.e., $\text{Var}[(r_1 + r_2)/2] = (\text{Var}[r_1] + \text{Var}[r_2])/4 = (\tau/dt + \tau/dt)/4 = \tau/(2dt)$, which will later permit a sensible continuum limit as $dt \to 0$ to yield a Markovian stochastic process \cite{Jacobs2006}.
The simplest Kraus operator for such a Gaussian measurement is $M_r = \sqrt{P(r|1)}\ket{1}\!\bra{1} + \sqrt{P(r|{-1})}\ket{0}\!\bra{0}$, which may be written in a more compact form as a function of the operator $\sigma_z$,
\begin{equation}
\label{eq:gausskraus}
M_r = \left( \frac{d t}{2 \pi \tau }\right)^{1/4} \exp\left[-\frac{(r - \sigma_z)^2 d t}{4 \tau} \right]. 
\end{equation}
Despite its simplicity, this Gaussian model is a reasonable approximation for a variety of experimental situations, including double-quantum-dot measurements with a quantum point contact \cite{Korotkov2001}, and superconducting transmon measurements with microwave resonators \cite{Korotkov2014}. 

Notably, the probability distribution $P(r|\rho)$ for causally obtaining a future $r$ from the current state $\rho$ may be conveniently expanded in terms of the single expectation value $z \equiv \tr{\rho\,\sigma_z}$ as
\begin{align}
\label{eq:predictivedistribution}
P(r|\rho) &= P(+1|\rho)\, G_1(r) + P(-1|\rho)\, G_{-1}(r) \nonumber \\
&= \frac{\left( 1 + z \right)}{2}\, G_{1}(r)   + \frac{\left( 1 - z \right) }{2}\, G_{-1}(r) ,
\end{align}
which allows all moments of $r$ to be easily calculated. For example, the first three moments are:
\begin{align}
\label{eq:momentspredictive}
\langle r \rangle = z, \quad \ \  \langle r^2 \rangle = 1 + \frac{\tau}{d t}, \ \ \quad \langle r^3 \rangle = \left( 1 + 3 \frac{\tau}{d t} \right) z.
\end{align}
All such moments for future $r$ are characterized solely by the expectation value $z$ of the measured observable in the qubit state $\rho$ immediately prior to the measurement. We will see in the next section that this feature will no longer be true for moments of past $r$.

Let us now consider a sequence of $N$ such generalized measurements $M_r$, with outcomes $r_j$, with $j = 1,\ldots,N$. Between each measurement, the qubit independently evolves for the time step $dt$ with Hamiltonian $H$, which we model by a separate unitary operator $U \equiv \exp(-i H dt/\hbar)$. 
The state of the qubit at the time $T = N\,dt$, given an initial state $\rho$ at time $t=0$ and the past set of outcomes $\vec{r} = (r_1, \ldots, r_N)$ is then:
\begin{align}
\label{eq:fullupdaterho}
	\rho_{\vec{r}} = \frac{ \big( M_{r_N} U \ldots M_{r_1} U \big) \rho \big( U^\dag M_{r_1}^\dag \ldots U^\dag M_{r_N}^\dag \big) }{P(\vec{r}|\rho) },
\end{align}
where the joint probability $P(\vec{r}|\rho) = \tr{E_{\vec{r}}\,\rho}$ of all measured results is governed by the positive operator from a joint POVM
\begin{align}
\label{eq:fullpovmop}
	E_{\vec{r}} = \big(U^\dag M_{r_1}^\dag \ldots U^\dag M_{r_N}^\dag\big) \big(M_{r_N} U \ldots M_{r_1} U\big).
\end{align}
This model describes the periodic monitoring of the observable $\sigma_z$ at the times $t_j = j\,dt$. In the continuum limit as $dt \to 0$ and $N\to\infty$, keeping $T = N\,dt$ constant, the unitary and measurement operators will commute up to second order in $dt$ such that each pair of operators $M_{r_j} U$ effectively describe the evolution within the same time step $[t_j, t_j+dt)$, and the evolution in Eq.~\eqref{eq:fullupdaterho} becomes equivalent to a stochastic master equation~\cite{Jacobs2006,Wiseman2009,Gambetta2008,Patti2017} that describes truly continuous-in-time observable monitoring. In what follows, however, we retain the explicitly discrete time steps $dt$ for numerical stability and conceptual clarity. The discrete model has the added benefit of also modeling physically discrete sequences of impulsive Gaussian measurements \cite{Diosi16,Haroche15}.

When the continuum limit $dt \to 0$ is taken, the widths of the Gaussian distributions $P(r|\pm 1)$ broaden and mostly overlap, so the distribution $P(r|\rho)$ in \eqref{eq:predictivedistribution} approximates a single Gaussian distribution centered at the expectation value $z(t) = \tr{\rho_{\vec{r}}\,\, \sigma_z}$:
\begin{align}
\label{eq:distforcontmeas}
	P(r(t)|\rho_{\vec{r}}) &\approx G_{z(t)}[r(t)] \\
	&= \sqrt{\frac{d t}{2 \pi \tau}} \exp{ \left( - \frac{\big[ r(t)-z(t) \big]^2 dt}{2 \tau} \right) }, \nonumber
\end{align}
where we have replaced the discrete index $j$ with the continuous time $t$.
It follows that in this limit the \emph{future} (uncollected) readout can itself be approximated as a moving mean Markovian stochastic process centered at the evolving expectation value $z(t)$
\begin{equation}\label{eq:prednoise}
r(t) = z(t) + \sqrt{\tau}\, \xi(t),
\end{equation}
where $\xi(t)$ is zero-mean additive white noise~\cite{Jacobs2006,Wiseman2009}, satisfying $\langle \xi(t) \xi(t') \rangle = \delta(t'-t)$. This understanding of the readout as an expectation value cloaked by additive noise is standard in the literature of continuous quantum measurements~\cite{Wiseman2009,
Diosi88,
Gambetta2008,KorotkovXYZ}, and has been applied with tremendous success in a variety of experiments~\cite{Murch2013,Weber2014,Foroozani2016}.

Note that the white noise expression in Eq.~\eqref{eq:prednoise} seems to give a simple prescription for how to learn information about the qubit evolution solely from the readout $r(t)$. For example, classical signal processing methods can reduce the zero-mean noise $\xi(t)$ and thus approximately recover the dynamics of the observable expectation value $z(t)$. This feature has been demonstrated for the observation of quantum jumps between $z = \pm 1$ \cite{Vijay2011,Slichter2016}. Moreover, by concurrently monitoring the three qubit Pauli operators $\sigma_x$, $\sigma_y$, and $\sigma_z$ and applying exponential filtering to the collected readouts, the dynamics of all three expectation values $x(t)$, $y(t)$ and $z(t)$ that determine the evolving qubit state $\rho(t)$ may be recovered simultaneously with reasonably high fidelity for individual measurement realizations~\cite{KorotkovXYZ}. This latter result is particularly startling, since it seems to challenge an interpretation of the quantum state $\rho$ as pertaining solely to an ensemble of realizations. 

The relation in Eq.~\eqref{eq:prednoise} is misleading, however, since it pertains only to an \emph{as-yet-uncollected} future readout, and does not yet describe the temporal structure of a readout that was collected in the past. Due to the informational backaction of the measurement, the readout and state evolution become temporally correlated, which effectively refines the distribution in Eq.~\eqref{eq:distforcontmeas} of the past readout and shifts the mean of Eq.~\eqref{eq:prednoise}. Strictly speaking, the relation in Eq.~\eqref{eq:prednoise} only holds at the final collected time of the readout, which still has an uncertain future.

\section{Observable dynamics from anterior and posterior measurements}
\label{sec:quantumsmoothing}

The previous section demonstrated that the future readout $r(t)$ is fully characterized by the expectation value $z(t)$ of the observable $\sigma_z$ with the causal qubit state $\rho(t)$. In this section we show that the past collected readout is not completely characterized by the causal qubit state, and derive a refined description of the implied dynamics of the measured observable that better agrees with the collected record.

\begin{figure}[t]
\includegraphics[width=0.9\columnwidth]{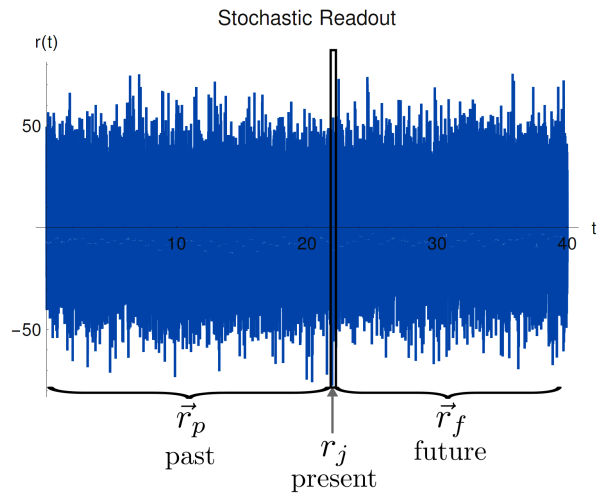}
	\caption{Partitioning of past collected readout. Given a temporal sequence of $N$ measurement results $\vec{r} = (r_1,\ldots,r_N)$ collected at times $(t_1,\ldots,t_N)$ separated by time steps $dt$, the best estimate of the monitored observable at an intermediate time $t_j$ depends on the distribution of the readout $r_j$ at that time, which depends on the future results $\vec{r}_f$ as well as the past history $\vec{r}_p$ because of the backaction of the measurement.}
\label{fig:pastandfuture}
\end{figure}

We now focus on what can be inferred about the qubit observable prior to a collected posterior record, as illustrated in Fig.~\ref{fig:pastandfuture}.  To do this, we partition the measured results $(\vec{r}_p, r,\vec{r}_f) = (r_1,\ldots,r_{j-1},r,r_{j+1},\ldots,r_N)$ into past results and future results relative to a particular past time $t_j$. We then derive the distribution $P(r|\vec{r}_p,\vec{r}_f,\rho)$ for the $r$ measured at $t_j$, conditioned not only on the past results $\vec{r}_p = (r_1, \ldots, r_{j-1})$ and initial state $\rho$, but also on the future results $\vec{r}_f = (r_{j+1}, \ldots, r_N)$ after $t_j$. As in Eq.~\eqref{eq:fullupdaterho}, the past results are fully encapsulated by the past causal state, 
\begin{equation}
	\rho_{\vec{r}_p} = \frac{ \big( U M_{r_{j-1}} U \ldots M_{r_1} U \big) \rho \big( U^\dag M_{r_1}^\dag \ldots U^\dag M_{r_{j-1}}^\dag U^\dag \big) }{P(\vec{r}_p|\rho) },
\end{equation}
Similarly, as in Eq.~\eqref{eq:fullpovmop}, the future results are fully represented by the future POVM element,
\begin{equation}
E_{\vec{r}_f} \equiv \big( U^\dag M_{r_{j+1}}^\dag \ldots U^\dag M_{r_N}^\dag \big) \big( M_{r_N} U \ldots M_{r_{j+1}} U \big).
\end{equation}
For ease of notation, we omit the time index $j$ for $r$, $E_{\vec{r}_f}$, and $\rho_{\vec{r}_p}$.  The need for both past and future quantities $(\rho_{\vec{r}_p},\,E_{\vec{r}_f})$ when describing the intermediate measurement result $r$ has been previously highlighted, with the pair dubbed a ``bidirectional quantum state'' in~\cite{Dressel13} and a ``past quantum state'' in~\cite{Molmer13}. These two quantities generalize the pure ``time-symmetric state'' $(\ket{\psi},\bra{\phi})$ from the ``two-vector formalism'' pioneered in \cite{Watanabe1955,Aharonov1964,AharonovWV}.

Applying Bayes' rule to the joint distribution $P(\vec{r}_p, r,\vec{r}_f|\rho)$ yields the desired distribution:
\begin{align}
\label{eq:bayes}
	P(r|\vec{r}_p ,\vec{r}_f, \rho) = \frac{P(\vec{r}_p,r,\vec{r}_f|\rho ) }{\int_{\mathbb{R}}P(\vec{r}_p,r,\vec{r}_f|\rho )\,dr }. 
\end{align}
From the preceding section, the joint distribution is:
\begin{equation}
\label{eq:probfuture}
P(\vec{r}_p, r,\vec{r}_f|\rho) = \tr{E_{\vec{r}_f} M_r \rho_{\vec{r}_p} M_r^\dag  }. 
\end{equation}
Combining Eqs.~\eqref{eq:gausskraus},~\eqref{eq:bayes} and~\eqref{eq:probfuture}
thus permits explicit calculation of the desired distribution:
\begin{align}
\label{eq:smootheddistribution}
&P(r|\vec{r}_p ,\vec{r}_f, \rho) = \frac{\Big( G_{1}(r) - G_{-1}(r) \Big) z_w}{\Big( 1 +  e^{-\tfrac{d t}{2 \tau} }  \Big) + \Big( 1 - e^{-\tfrac{d t}{2 \tau} }   \Big) z_c}  \nonumber \\
&\qquad + \frac{ \tfrac{1}{2} \Big( G_{1}(r) + G_{-1}(r) + 2 e^{-\tfrac{d t}{2 \tau} } G_0(r) \Big) }{\Big( 1 +  e^{-\tfrac{d t}{2 \tau} }  \Big) + \Big( 1 - e^{-\tfrac{d t}{2 \tau} }   \Big) z_c} \nonumber \\ 
&\qquad + \frac{  \tfrac{1}{2} \Big( G_{1}(r) + G_{-1}(r) - 2 e^{-\tfrac{d t}{2 \tau} } G_0(r) \Big) z_c }{\Big( 1 +  e^{-\tfrac{d t}{2 \tau} }  \Big) + \Big( 1 - e^{-\tfrac{d t}{2 \tau} }   \Big) z_c}, 
\end{align}
which depends on only two quantities containing the monitored observable:
\begin{align}\label{eq:wv}
z_w &\equiv \text{Re}\frac{\tr{E_{\vec{r}_f}\, \sigma_z\,\rho_{\vec{r}_p}}}{\tr{E_{\vec{r}_f}\, \rho_{\vec{r}_p}} }, &
z_c &\equiv \frac{ \tr{E_{\vec{r}_f}\, \sigma_z\, \rho_{\vec{r}_p}\, \sigma_z} }{ \tr{E_{\vec{r}_f}\, \rho_{\vec{r}_p}} }, 
\end{align}
neither of which are an expectation value. Instead, $z_w$ is the real part of a (generalized) \emph{weak value}~\cite{AharonovWV,Jozsa07,DresselWVreview,Dressel2015WVInt} in its role as a first-order conditioned expectation value~\cite{DresselWVreview,Dressel2015WVInt}, and $z_c$ is a second-order contribution \cite{DiLorenzo12}. Due to the assumption of Gaussian statistics, these first two orders are sufficient to fully characterize the distribution.

For comparison with Eq.~\eqref{eq:momentspredictive}, the first three moments of the distribution $P(r|\vec{r}_p,\vec{r}_f,\rho)$ are: 
\begin{subequations}
\begin{align}
\label{eq:firstmomentsmoothed}
\langle r \rangle_S &= z_S,  \\
	\langle r^2 \rangle_S &= \frac{\tau}{dt} + \frac{\tfrac{1}{2}(1 + z_c)}{\tfrac{1}{2}\Big( 1 +  e^{-\tfrac{dt}{2 \tau} }  \Big) + \tfrac{1}{2}\Big( 1 - e^{-\tfrac{dt}{2 \tau} }   \Big) z_c},  \\
\langle r^3 \rangle_S &= \left( 1 + 3 \frac{\tau}{dt} \right) z_S.
\end{align}
\end{subequations}
Remarkably, after taking into account subsequent measurement outcomes the mean of the intermediate $r$ shifts to a refined (smoothed) estimate $z_S$ instead of the traditionally accepted expectation value $z$ that we obtained in the previous section (compare with Eq.~\eqref{eq:momentspredictive}). 
This \emph{smoothed estimate} of $\sigma_z$ is the central quantity of this paper, 
\begin{align}
\label{eq:smoothedZ}
z_S \equiv \frac{z_w}{ \tfrac{1}{2} \Big( 1 +  e^{-\tfrac{dt}{2 \tau} }  \Big) + \tfrac{1}{2} \Big( 1 - e^{-\tfrac{dt}{2 \tau} }   \Big) z_c},
\end{align}
and depends upon both $z_w$ and $z_c$. 

Note that for weak measurements with large variance, $\tau/dt \gg 1$, the second-order contribution $z_c$ is suppressed, and the smoothed estimate converges to the first-order weak value $z_w$. The smoothed estimate therefore inherits the behaviour of the weak value, to a degree that depends on the coarseness of the time steps. For example, the smoothed estimate can take values outside the range $[-1,1]$ of possible observable values for $\sigma_z$~\cite{Diosi06,Diosi16}, as shown in the next section. Importantly, \emph{this convergence to the weak value becomes exact in the continuum limit as $dt \to 0$.}
Moreover, the continuum limit of the full distribution in Eq.~\eqref{eq:smootheddistribution} is a single Gaussian distribution similar to Eq.~\eqref{eq:distforcontmeas} but centered on the \emph{smoothed estimate} $z_S$ of $\sigma_z$, which in turn converges to $z_w$:
\begin{align}
\label{eq:distforcontmeasSmoothed}
	P\left( r(t)|\vec{r}_p,\vec{r}_f,\rho \right) &\approx G_{z_w(t)}[r(t)] \\
	&= \sqrt{\frac{dt}{2 \pi \tau}} \exp{ \left( - \frac{\big[ r(t)-z_w(t) \big]^2 dt}{2 \tau} \right) }. \nonumber
\end{align}
Therefore, arguments identical to the preceding section imply that the \emph{past} (already collected) continuous readout still has the structure of a moving mean stochastic process, but instead following a \emph{weak value} of $\sigma_z$:
\begin{equation}\label{eq:noisewv}
r(t) = z_w(t) + \sqrt{\tau}\, \xi(t).
\end{equation}

This key result implies that using classical signal processing techniques to reduce the zero-mean white noise $\xi(t)$ on a collected readout $r(t)$ will \emph{not} recover the expectation value $z(t)$ as might be expected from the previous section. Instead, such techniques will recover the smoothed (weak-valued) estimate $z_S(t)\to z_w(t)$ that properly takes into account the temporal correlations in the signal caused by the measurement backaction. This discrepancy explains the limited fidelity of the state reconstruction seen in \cite{KorotkovXYZ} when exponentially filtering simultaneous observable readouts. The reasonably high fidelities that were still obtained are explained by the fact that the smoothed estimate $z_w(t)$ can often remain close to the expectation value, as shown in the next section.

\section{A better description of past observable dynamics}
\label{sec:quantumsmoothingfigures}

The previous section established that a smoothed observable estimate more closely describes the observed readout than an expectation value. In this section, we numerically simulate an explicit example of a monitored qubit Rabi oscillation to demonstrate the practical significance of this result. Specifically, we define two figures of merit that contrast an expectation value with a smoothed estimate and show that the smoothed estimate systematically outperforms the expectation value. Importantly, we consider individual measurement realizations, not ensemble averages.

\begin{figure}[t]
    \includegraphics[width=\columnwidth]{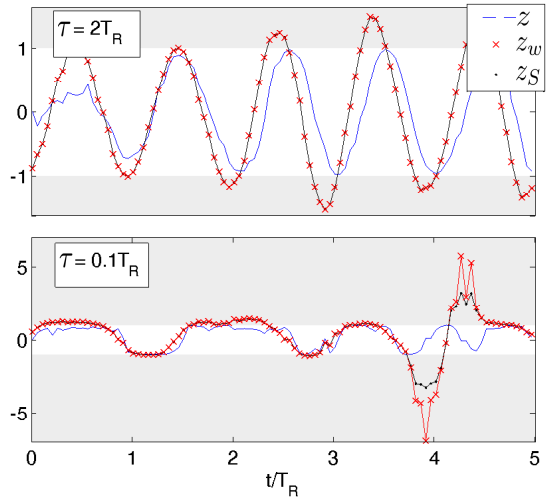}
	\caption{Evolution of observable estimates. Shown are single realizations of qubit Rabi oscillations in the $x$-$z$ plane with period $T_R$ while monitoring $\sigma_z$ with collapse timescale $\tau$ at periodic time steps $dt$. Compared are the expectation value $z$ (blue, solid), smoothed value $z_S$ (black, dotted), and weak value $z_w$ (red, hashed). The upper plot shows a weaker measurement regime with $\tau/T_R = 2$, and with $dt/\tau = 1/40$ so $z_S \approx z_w$. The lower plot shows a stronger measurement regime with $\tau/T_R = 1/10$, and with $dt/\tau = 1/2$, so $z_S$ deviates from sufficiently large values of $z_w$.
    }
\label{fig:evolSmoothed}
\end{figure}

Consider the periodic monitoring of $\sigma_z$ at time steps $dt$ for a total duration $T$, with characteristic collapse timescale $\tau$, on a qubit driven by a Hamiltonian 
\begin{equation}
H = \hbar\,\Omega\, \frac{\sigma_y}{2}, 
\end{equation}
where the Pauli matrix $\sigma_y$ generates Rabi oscillations in the $x$-$z$ plane, and $T_R \equiv 2\pi/\Omega$ is the period of these oscillations. Figure~\ref{fig:evolSmoothed} illustrates the distinction between the expectation value $z$ (blue, solid), the smoothed value $z_S$ (black, dotted), and the weak value $z_w$ (red, hashed) for such a monitored oscillation. In both plots $dt/T_R = 1/20$ is held constant, with the upper plot showing a weaker measurement with $\tau/T_R = 2$, and the bottom plot showing a stronger measurement with $\tau/T_R = 1/10$. The weaker measurement in the upper plot exhibits noisy Rabi oscillations since the dynamics are only weakly perturbed by the monitoring. For this upper plot, $dt/\tau = (1/20)/2 = 1/40$, so the smoothed estimate $z_S$ and weak value $z_w$ are essentially indistinguishable. The stronger measurement in the lower plot exhibits quantum jumps between measurement eigenstates, to which the qubit is pinned by the quantum Zeno effect \cite{Vijay2011,Slichter2016}. For this lower plot, $dt/\tau = (1/20)/(1/10) = 1/2$, so the smoothed estimate $z_S$ is visibly distinct from the weak value $z_w$ when the latter becomes sufficiently large. Note that the weak value $z_w$ can exceed the eigenvalue range of $[-1,1]$ when the readout $r(t)$ is statistically unlikely.

To answer the question whether the expectation value $z$ or the smoothed value $z_S$ better follows a single readout realization $r$ quantitatively, we establish two figures of merit. First, we consider the mean squared error between the $N$-dimensional readout vector $\vec{r}$ for all $N$ time steps $dt$, and the dynamical estimate vectors $\vec{z}$ and $\vec{z_S}$. The mean squared error is defined for any two vectors $\vec{v}$ and $\vec{w}$ of length $N$ as $\textnormal{MSE}(\vec{v},\vec{w}) = \sum_{j=1}^N (v_j - w_j)^2/N$.
Notably, the mean squared error is the primary figure of merit used in classical filtering and estimation theory~\cite{filteringbook} to find optimal estimates for the ``true'' value of a noise-polluted signal. 
We treat the readout $\vec{r}$ as such a noise-polluted signal, and define the relative mean squared error,
\begin{equation}
\label{eq:relmse}
Q(z_S,z) \equiv \frac{\textnormal{MSE}(\vec{r},\vec{z}) - \textnormal{MSE}(\vec{r},\vec{z}_S) }{\textnormal{MSE}(\vec{r},\vec{z}_S) },
\end{equation}
such that $Q(z_S,z) > 0$ if and only if the smoothed estimate $\vec{z}_S$ is a better fit to the measured readout $\vec{r}$ than the expectation value $\vec{z}$. 

The second figure of merit we adopt is a hypothesis test. Namely, let us assume prior probabilities $P(z)$ and $P(z_S)$ for models in which the readout $\vec{r}$ fits $\vec{z}$ or $\vec{z}_S$ respectively.
Bayes' rule allows us to express the probability of the estimate $\vec{z}$ given the observed measurement record $\vec{r}$ as $P(\vec{z}|\vec{r}) = P(\vec{r}|\vec{z}) P(z)/ P(\vec{r})$. Similarly, $P(\vec{z}_S|\vec{r}) = P(\vec{r}|\vec{z}_S) P(z_S)/P(\vec{r})$. Assuming equal prior probabilities for both hypotheses, $P(z) = P(z_S)$, we can then define the hypothesis test ratio as
\begin{align}
\label{eq:factorR}
R(z_S,z) \equiv \frac{  P(\vec{z}_S|\vec{r})}{  P(\vec{z}|\vec{r})} = \frac{P(\vec{r}|\vec{z}_S)  P(z_S)}{P(\vec{r}|\vec{z})  P(z)} = \frac{P(\vec{r}|\vec{z}_S) }{P(\vec{r}|\vec{z})  },
\end{align}
where $P(\vec{r}|\vec{z}) = \prod_j P(r_j| \rho_{\vec{r}_p})$ and $P(\vec{r}|\vec{z}_S) = \prod_j P(r_j| \rho_{\vec{r}_p},E_{\vec{r}_f})$ can be calculated from Eqs.~\eqref{eq:predictivedistribution} and~\eqref{eq:smootheddistribution}.
The ratio $R$ discriminates the likelihood of the estimates $z$ or $z_S$, given the observed record $\vec{r}$.
We use its natural logarithm as a figure of merit to decide between the two alternatives: that is, $\ln R > 0$ if and only if the smoothed estimate is more probable than the expectation value. 

\begin{figure}[t]
\begin{center}
	\includegraphics[width=\columnwidth]{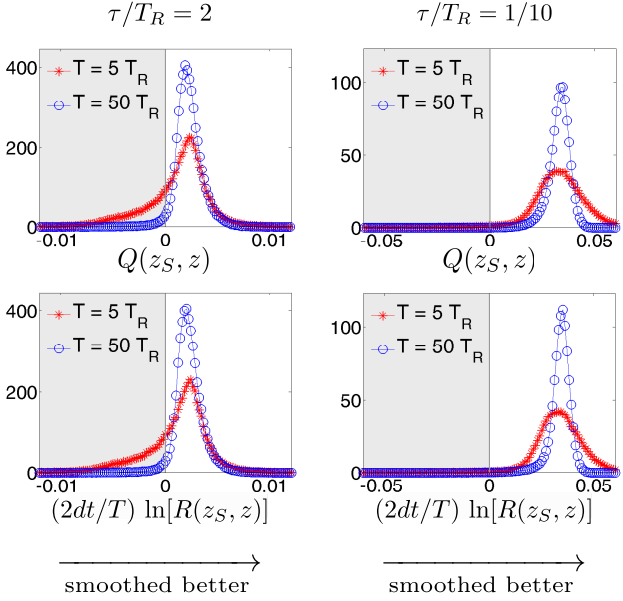}
\end{center}
	\vspace{-1em}
	\caption{Estimate comparisons. Normalized histograms comparing the expectation value $z$ to the smoothed value $z_S$, for $10^5$ realizations monitoring $\sigma_z$ of a qubit oscillating with Rabi period $T_R$ for a duration $T$, with time steps $dt = T_R/100$. (red, hashed) Short duration $T/T_R = 5$. (blue, circle) Long duration $T/T_R = 50$. (top row) Relative mean squared error $Q(z_S,z)$, see Eq.~\eqref{eq:relmse}. (bottom row) Hypothesis test log-ratio $\ln\left[R(z_S,z)\right]$, see Eq.~\eqref{eq:factorR}, scaled by $2dt/T$ to show correspondence to $Q(z_S,z)$. (left column) Weaker measurement regime with $\tau/T_R = 2$. (right column) Stronger measurement regime with $\tau/T_R = 1/10$. 
For both figures of merit, positive values indicate that $z_S$ is a better estimate than $z$. The fraction of realizations for which $Q(z_S,z) > 0$ (or $\ln\left[R(z_S,z)\right] > 0$) for $T/T_R = \{5,50\}$ is higher than $\{0.735,0.968\}$ for weaker measurements, and higher than $\{0.976,0.999\}$ for stronger measurements, respectively. 
}
\label{fig:testingSmoothed}
\end{figure}

Figure~\ref{fig:testingSmoothed} shows histograms of both the relative mean squared error $Q(z_S,z)$ (top row) and the hypothesis test log-ratio $\ln[R(z_S,z)]$ (bottom row), computed for $10^5$ realizations using fixed time steps of $dt/T_R = 1/100$ to approximate the continuum limit in two regimes: (left column) a weaker regime with $\tau/T_R = 2$, and (right column) a stronger regime with $\tau/T_R = 1/10$. For longer trajectories with $T/T_R = 50$ (blue circles), the smoothed estimate is better (i.e., has a positive discriminator) $96.8\%$ of the time in the weaker regime and $99.9\%$ of the time in the stronger regime. Even for relatively short trajectories with $T/T_R = 5$, the smoothed estimate is better $73.5\%$ of the time in the weaker regime and $97.6\%$ of the time in the stronger regime. The improvement in the estimate is approximately linear in the inverse collapse timescale $\tau^{-1}$, since the $T/\tau$ determines the signal-to-noise ratio of the stochastic readout. The factor of $20$ improvement in performance expected between the weaker and stronger regimes is confirmed by the shift in mean of the histograms in Figure~\ref{fig:testingSmoothed}---note that the means do not shift with the duration $T$, since the plotted discriminators are effectively normalized per unit $T$. These results confirm that one should consider the collected readout to better follow the smoothed estimate $z_S(t) \approx z_w(t)$ given in Eq.~\eqref{eq:wv}, \emph{not} the expectation value as naively expected from Eq.~\eqref{eq:prednoise}. 

Observe that in Figure~\ref{fig:testingSmoothed}, we have scaled the hypothesis test log-ratio $\ln[R(z_S,z)]$ by a factor $(2dt/T)$ to make its correspondence to the relative mean squared error $Q(z_S,z)$ evident, and to serve as a consistency check for the simulations. This correspondence may be explained by noting that in the time-continuous limit the probabilities $P(\vec{r}|\vec{z}) = \prod_j P(r_j| \rho_{\vec{r}_p})$ and $P(\vec{r}|\vec{z}_S) = \prod_j P(r_j| \rho_{\vec{r}_p},E_{\vec{r}_f})$ may be approximated by products of Gaussian distributions, as in Eqs.~\eqref{eq:distforcontmeas} and~\eqref{eq:distforcontmeasSmoothed}, respectively. It follows that the hypothesis test simplifies, 
\begin{align}
\frac{2dt}{T}\,\ln\left[ R(z_S,z) \right] \approx 
 \frac{dt \textnormal{MSE}(\vec{r},\vec{z}_S)}{\tau} Q(z_S,z) \approx Q(z_S,z),
\end{align}
where in the last step we have used that in the continuum limit
$\textnormal{MSE}(\vec r,\vec z_S) = \sum_j (r_j-z_{S,j})^2 /N \approx \tau \sum_j (\xi_j)^2 /N = \tau/dt$,
from Eq.~\eqref{eq:distforcontmeasSmoothed} and the fact that the white noise $\xi$ at any time step has variance $1/dt$. This relationship between $Q(z_S,z)$ and $\ln\left[R(z_S,z)\right]$ in the time-continuous limit is correctly confirmed in Fig.~\ref{fig:testingSmoothed}. Importantly, this numerical equivalence between the two \emph{a priori} distinct figures of merit confirms the white noise relation in Eq.~\eqref{eq:noisewv}, and thus that $z_S \approx z_w$ is in fact the \emph{minimum} mean squared error estimate for individual realizations of the readout $r(t)$.

\section{Smoothed estimates by an ignorant third party}
\label{sec:smoothingphysical}

Crucially, the smoothed observable estimate derived in the preceding sections is not merely an artificial best fit to a past record, but is also a \emph{predictive} quantity with operational meaning that extends beyond the original collected record. To see this, we now consider a situation where two observers monitor different observables on the same system. The task at hand will be for the first observer to estimate what was measured by the second observer. For this task, we now show that a smoothed estimate using partial information is not only operationally better than an expectation value that uses partial information, but can even be better than an expectation value that uses all available information.

For specificity, consider an agent $\mathcal{Z}$ who monitors $\sigma_z$ on a qubit, as described in the previous sections, while a second agent $\mathcal{X}$ simultaneously monitors the distinct observable $\sigma_x$ in a similar way (as considered in Refs.~\cite{KorotkovXYZ,XZHacohen-Gourgy16,LPGPJD16,Perarnau16}). 
We assume characteristic collapse timescales $\tau_z$ and $\tau_x$ for Gaussian Kraus operators $M_{r_z}$ and $N_{r_x}$ measuring $\sigma_z$ and $\sigma_x$, respectively, with $\tau_z < \tau_x$ so that the agent $\mathcal{Z}$ causes the majority of the measurement backaction. After both $\mathcal{X}$ and $\mathcal{Z}$ monitor
for a duration $T = N\,dt$, 
they each possess one measurement record, $\vec{r}_z$ or $\vec{r}_x$. We now consider two distinct scenarios: (A) An omniscient third agent $\mathcal{O}$ examines both measurement records and estimates both $\sigma_z$ and $\sigma_x$ using all information, and (B) The agent $\mathcal{Z}$ uses only the record $\vec{r}_z$ to estimate $\sigma_x$ without knowledge of what agent $\mathcal{X}$ actually measured.

\begin{figure*}[t]
\begin{center}
  \includegraphics[width=\textwidth]{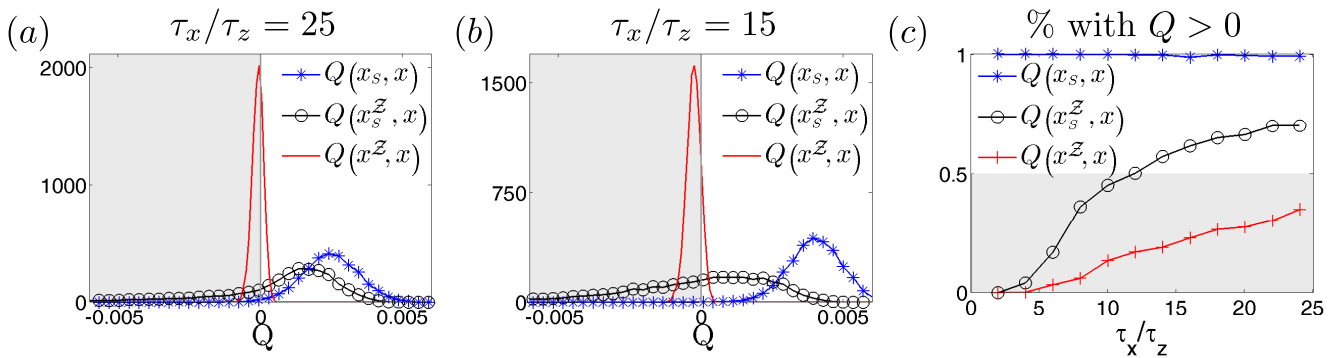}
\end{center}
	\vspace{-2em}
\caption{Informationally incomplete estimate comparisons. Agents $\mathcal{Z}$ and $\mathcal{X}$ monitor $\sigma_z$ and $\sigma_x$, respectively, of a qubit oscillating with Rabi period $T_R$ for a duration $T/T_R = 100$, with fixed time steps $dt/T_R = 1/100$, and collapse timescale $\tau_z/T_R = 1/10$.
	(a), (b) Normalized histograms with $10^4$ realizations, comparing the mean squared errors $Q$ relative to the reference expectation value $x$ that uses all information contained in the causal quantum state evolving from both measurement records $r_z$ and $r_x$. The ignorant expectation value $x^{\mathcal{Z}}$ (red, solid) and ignorant smoothed estimate $x_S^{\mathcal{Z}}$ (black, circles) use incomplete information contained only in the record $r_z$, while the optimal smoothed estimate $x_S$ (blue, starred) also uses the information in $r_x$. For weaker $\mathcal{X}$ monitoring (a), with $\tau_x/\tau_z = 25$, the ignorant smoothed estimate $x_S^{\mathcal{Z}}$ still beats $x$ in more than $72\%$ of the realizations,
while the best smoothed estimate $x_S$ beats $x$ in more than $98\%$ of the realizations, and the ignorant expectation value $x^{\mathcal{Z}}$ beats $x$ in only $32\%$ of the realizations.
For stronger $\mathcal{X}$ monitoring (b), with $\tau_x/\tau_z = 15$, the advantage of the ignorant smoothed estimate $x_S^\mathcal{Z}$ is reduced, with only $54\%$ of the realizations beating $x$.
(c) Relative mean square error $Q$ as a function of the relative monitoring strength $\tau_x/\tau_z$, showing that for $\tau_x/\tau_z \gg 1$ the ignorant smoothed estimate $x_S^{\mathcal{Z}}$ reliably outperforms the expectation value $x$, despite having restricted information.
}
\label{fig:smoothedXZ}
\end{figure*}

For the omniscient observer $\mathcal{O}$ in scenario (A), the access to both sets of outcomes one allows the derivation of smoothed estimates $z_S$ and $x_S$ precisely as in Section~\ref{sec:quantumsmoothing} from the joint probability of obtaining $r_x$ and $r_z$ conditioned on both past and future outcomes $P\big( r_x,r_z |\vec{r}_{z,p},\vec{r}_{z,f}, \vec{r}_{x,p},\vec{r}_{x,f}, \rho \big)$. The form of the smoothed estimates is as in Eqs.~\eqref{eq:smoothedZ} and \eqref{eq:wv}, but with a modified bidirectional state consisting of
\begin{equation}
\rho_{\vec{r}_{z,p},\vec{r}_{x,p}} \equiv \frac{ \big( U N_{r_{x,j-1}}M_{r_{z,j-1}} U \ldots N_{r_{x,1}} M_{r_{z,1}} U \big) \rho \ldots \big) }{P(\vec{r}_{z,p},\vec{r}_{x,p}|\rho) },
\end{equation}
and
\begin{equation}
	E_{\vec{r}_{z,f},\vec{r}_{x,f}} \equiv \big( U^\dag M_{r_{z,j+1}}^\dag N_{r_{x,j+1}}^\dag \ldots U^\dag M_{r_{z,N}}^\dag N_{r_{x,N}}^\dag \big) \big( \ldots \big).
\end{equation}
Note that our model here interleaves the measurements of $\sigma_x$ and $\sigma_z$ for simplicity; in the continuum limit, $dt \to 0$, the measurements become effectively simultaneous \cite{LPGPJD16,XZHacohen-Gourgy16}. As in the last section, the smoothed estimates obtained in this way fit the measurement output better than the expectation values $x = \tr{\rho_{\vec{r}_{z,p},\vec{r}_{x,p}}\, \sigma_x}$ and $z = \tr{\rho_{\vec{r}_{z,p},\vec{r}_{x,p}}\, \sigma_z}$ obtained from the most informationally complete causal state of the qubit. 

The more interesting case is scenario (B), where agent $\mathcal{Z}$ has incomplete information from which to construct an estimate. Let $x_S^{\mathcal{Z}}$ be the smoothed estimate of $\sigma_x$ based on the bidirectional state known to $\mathcal{Z}$, which takes into account only the measurement collapses from the monitoring of $\sigma_z$:
\begin{align}
	\rho_{\vec{r}_{z,p}} &\equiv \frac{ \big( U M_{r_{z,j-1}} U \ldots M_{r_{z,1}} U \big) \rho \big( \ldots \big) }{P(\vec{r}_{z,p}|\rho) }, \\
E_{\vec{r}_{z,f}} &\equiv \big( U^\dag M_{r_{z,j+1}}^\dag \ldots U^\dag M_{r_{z,N}}^\dag \big) \big( \ldots \big).
\end{align}
Similarly, let $x^{\mathcal{Z}}$ be the expectation value of $\sigma_x$ based on the causal state $\rho_{\vec{r}_{z,p}}$ known to $\mathcal{Z}$. How good are the two ignorant estimates $x^{\mathcal{Z}}$ and $x_S^{\mathcal{Z}}$ compared to those made by the omniscient observer $\mathcal{O}$?

Using the omniscient expectation value $x$ as a reference, a fixed long duration $T/T_R = 100$, fixed time steps $dt/T_R = 1/100$, and fixed collapse timescale $\tau_z/T_R = 1/10$, Figure~\ref{fig:smoothedXZ} shows the relative mean squared error for the ignorant expectation value, $Q(x^{\mathcal{Z}},x)$ (red, solid), the ignorant smoothed estimate $Q(x_S^{\mathcal{Z}},x)$ (black, circles), and the omniscient smoothed estimate $Q(x_S,x)$ (blue, starred). This latter quantity shows the maximum improvement for reference, with $\sim$99\% of realizations consistently favoring the omniscient smoothed estimate. The left plot (a) shows the case when the monitoring of $\mathcal{X}$ is substantially weaker than $\mathcal{Z}$, $\tau_x/\tau_z = 25$, so perturbs the evolution less in comparison. The middle plot (b) shows slightly less weak monitoring by $\mathcal{X}$, $\tau_x/\tau_z = 15$. The right plot (c) shows the fraction of cases that are better than the omniscient expectation value $x$ (i.e., where $Q>0$) as the ratio between monitoring strengths $\tau_x/\tau_z$ varies. Unsurprisingly, the ignorant expectation value $x^{\mathcal{Z}}$ is always worse on average than the omniscient expectation value $x$ because of the loss of information. Surprisingly, however, the ignorant smoothed estimate $x_S^{\mathcal{Z}}$ can \emph{outperform} the omniscient expectation value for predicting what the agent $\mathcal{X}$ actually measured when $\tau_x/\tau_z \gg 1$. In (a), more than $72\%$ of the realizations favor the ignorant smoothed estimate, even though the monitoring of $\sigma_x$ produces backaction that is \emph{not} negligible, with $T/\tau_x = 40$. In (b), more than $54\%$ of the realizations favor the ignorant smoothed estimate, showing a reduction in advantage as the sources of backaction become comparable. In the opposite limit as $\tau_x/\tau_z \to \infty$, the monitoring by $\mathcal{X}$ no longer perturbs the system and the ignorant estimates converge to the omniscient estimates, $x^{\mathcal{Z}} \to x$ and $x_S^{\mathcal{Z}} \to x_S$.

To emphasize the significance of this result, we note that a similar situation to scenario (B) has been discussed by Guevara and Wiseman \cite{GuevaraWiseman15}, who conclude that using an entire collected measurement record allows one to construct a ``smoothed state'' $\rho_S$ that is closer in fidelity to the (typically pure) maximally informative state $\rho_{\vec{r}_{z,p},\vec{r}_{x,p}}$ known to an omniscient observer $\mathcal{O}$ than the (mixed) state $\rho_{\vec{r}_{z,p}}$ constructed from the incomplete information known to $\mathcal{Z}$. The surprising extension to this result that we show here is that the informationally \emph{incomplete} smoothed estimate $x_S^{\mathcal{Z}}$ can outperform even the best expectation value $x$ known to the omniscient observer $\mathcal{O}$. Evidently, the omniscient state $\rho_{\vec{r}_{z,p},\vec{r}_{x,p}}$ is not informationally complete when it comes to the content of the collected readout.

\section{Discussion}
\label{sec:conclusion}

We have shown that, contrary to traditional wisdom, the collected readout of a continuous quantum measurement is not centered on the expectation value of the monitored observable. Instead, the readout is centered on a modified moving mean, which is a smoothed observable estimate that converges to the real part of a generalized weak value in the time-continuous limit.  The physical reason that the accumulated data follows this smoothed estimate, as opposed to the expectation value used to causally generate the same data, is that the partial measurement collapses create nontrivial correlations between past and future measurements that are only exposed in retrospect. The smoothed observable estimate provides an objectively better description of the readout than what can be accounted for solely from knowledge of the causal state of the qubit. Notably, this correspondence applies to single measurement realizations, without the need for ensemble averages, and without the need for additional postselection. Importantly, this result implies that applying classical signal processing techniques to the measurement output will not reveal information about the causal state of the system, but rather information about the smoothed estimate of the monitored observable, which bounds the fidelity of any state tomography scheme based on classical signal processing of the readout.

We have also shown that the smoothed estimate from the readout has operational meaning beyond the scope of a single measured observable. That is, an agent with access only to their own measurement record can still construct a meaningful smoothed estimate for a second observable being concurrently monitored by a second agent. Provided the second measurement is sufficiently weak compared to the first measurement, this informationally incomplete smoothed estimate will still be more consistent with the experimental output of the second agent than any quantity derived from the informationally complete causal qubit state. This observed improvement over the best causal quantum state estimate could have interesting applications for experimental parameter estimation and model verification in situations where one only has partial access or incomplete information about a system, which remains to be investigated.

\begin{acknowledgments}
We thank Alexander Korotkov and Howard Wiseman for helpful discussions. This work was supported by U.S. Army Research Office Grant No. W911NF-15-1-0496. We also acknowledge partial support by the Perimeter Institute for Theoretical Physics. Research at Perimeter Institute is supported by the Government of Canada through Industry Canada and by the Province of Ontario through the Ministry of Economic Development and Innovation.
\end{acknowledgments}

%

\end{document}